\documentclass[aps,prl,amsmath,amssymb,reprint,groupedaddress]{revtex4-1}
\usepackage{hyperref}
\usepackage{graphicx}
\usepackage{subfigure} 
\usepackage{dcolumn} 
\usepackage{bm}
\usepackage{mathtools}
\usepackage{float}
\usepackage{ulem}

\usepackage{soul}
\usepackage{color}
\begin{document}

\title{First-passage time distribution for random walks on complex networks using inverse Laplace transform and mean-field approximation}
\author{Mu Cong Ding}
\affiliation{Department of Computer Science and Engineering,\\The Hong Kong University of Science and Technology}
\author{Kwok Yip Szeto*}
\affiliation{Department of Physics,\\The Hong Kong University of Science and Technology}

\begin{abstract}
We obtain an exact formula for the first-passage time probability distribution for random walks on complex networks using inverse Laplace transform. We write the formula as the summation of finitely many terms with different frequencies corresponding to the poles of Laplace transformed function and separate the short-term and long-term behavior of the first-passage process. We give a formula of the decay rate $\beta$, which is inversely proportional to the characteristic relaxation time $\tau$ of the target node. This exact formula for the first-passage probability between two nodes at a given time can be approximately solved in the mean field approximation by estimation of the characteristic relaxation time $\tau$. Our theoretical results compare well with numerical simulation on artificial as well as real networks.
\end{abstract}

\maketitle
\section{I. INTRODUCTION}
Random walks on complex media have been extensively studied in recent years because of its central role in analyzing real-world stochastic processes\cite{Douglas1995}. The first-passage process of random walks is one of the main focus in this field because of its fundamental importance in stochastic processes that are triggered by first-passage event\cite{redner2001guide,lovasz1993random}. Different kinds of geometries have been considered as the media, such as regular lattice\cite{lawler2010random}, fractals\cite{ben2000diffusion,song2006origins}, self-similar networks\cite{PhysRevE.85.026113,0305-4470-22-7-019}, scale-free networks\cite{song2005self,pu2010mixing}, and as LÃ©vy flights\cite{koren2007first,PhysRevLett.99.148701}. Among them, random walks on networks not only serves as a useful tool for revealing the underlying characteristics of structures\cite{PhysRevE.75.016102}, but also provide a general model for the study of target searching\cite{PhysRevE.64.046135, PhysRevLett.89.248701}, information spreading\cite{PhysRevLett.86.3200,barrat2008dynamical} and diffusion\cite{Metzler20001}. 
The mean first passage time (MFPT) is a vital scale for random walk on networks and has been well studied in the past decade\cite{PhysRevLett.92.118701}. Comparing to the MFPT, the first-passage probability distribution gives more detailed information about the characteristics of the initial node, target node, and overall network structures. We are going to derive an exact formula for the first-passage probability at a given time, which separate terms with different time dependence and reveal the changing behavior of the first-passage process as a function of time. From this point of view, we will link the short-term and long-term behavior with different parameters of the topological structure of the network. Our result will serve as a useful guide for not only the characterization of the network but also the optimization of the first-passage process.

\section{II. THE FORMALISM OF FIRST PASSAGE TIME DISTRIBUTION USING LAPLACE TRANSFORM}
We consider a connected and undirected finite network of N nodes which can be represented by its adjacency matrix $A_{ij}$, with its element 
$A_{ij}= A_{ji}=1$  if the nodes $i,j$ are connected, and is zero otherwise. 
The degree of a node $i$ , denoted by $k_i$, is the number of connected neighbors, which is given by $k_i=\sum_{j}A_{ij}$. 
If there are $N_k$ nodes of degree $k$ in the network, then the probability of finding a node with degree $k$ is  $P_k={N_k}/N$ and the average degree of the network is $\left<k\right>=\sum_{k}k P_k$.

The random walk on this network is a discrete-time stochastic process. The walker at node $i$ and time $t$ selects one of its $k_i $ neighbors randomly at time $t+1$ for his next move, implying that the transition probability for the walker to go from node $i$ to node $j$ is given by $S_{ij}=A_{ij}/k_i$. Let us denote the probability of finding the random walker initially at node $i$, to reach node $j$ at time $t>0$ by $P_{ij}(t)$.  Note that the initial probability distribution is $P_{ij}^{0}=\delta_{ij}$. 
The evolution of this probability distribution is governed by the following master equation, 
$P_{ij}(t+1)=\sum_{\nu\in\text{Nei}(j)} P_{i \nu}(t) S_{\nu j}$, where $\text{Nei}(j)$ represents the set of neighbors of node $j$. 
By iterating this expression for a random walker, we can rewrite the evolution equation in matrix form by representing the probability distribution at time $t$ as a row vector $\mathbf{P}(t)$,
\begin{equation}
\mathbf{P}(t)=\mathbf{P}(t-1)\mathbf{S}=\mathbf{P^0_i}\mathbf{S}^{t}
\end{equation}
where $\mathbf{P^0_i}$ with the matrix elements $P_{ij}^{0}$ indicates that the initial node is $i$. We have the symmetric relation, $k_i P_{ij}(t)=k_j P_{ji}(t)$ due to the symmetric matrix $A_{ij}$. We also note that the stationary probability distribution $P_{ij}^\infty=k_j/(\sum_k kN_k)=k_j/(N\left<k\right>)$ only depends on the target node $j$.
 
We next introduce the auxiliary function $Q_{ij}(t)=P_{ij}(t)-P_{j}^\infty$, which converges to zero at the infinite time limit, as the random walk process eventually reaches the stationary probability distribution. 
From the evolution equation of $ P_{ij}(t)$, we get the evolution equation for $Q_{ij}(t)$ as 
\begin{equation}
Q_{ij}(t)=P_{ij}(t)-P_{j}^\infty=\left(\mathbf{P^0_i}-\mathbf{P^\infty}\right)\mathbf{S}^{t}\mathbf{e_{j}}
\end{equation}
where $\mathbf{e_{j}}$ is the column unit vector with the $j$-th entry equals to $1$. 
Note that $\widetilde{P_{ij}}(z)=\sum_{t=0}^{\infty}z^{-t}P_{ij}(t)$ converges only when $|z|>1$ but this problem is resolved with the auxiliary function $Q_{ij}(t)$ when the stationary probability $ P_{j}^\infty $ has been subtracted from $ P_{ij}(t)$. 
Since  $Q_{ij}(t)$ depends on the initial node $i$ and target node $j$,  and involves the transition probability matrix $\mathbf{S}$, 
it is related to the structure of the network. 
The Laplace transform of $Q_{ij}(t)$ is $\widetilde{Q_{ij}}(z)= \sum_{t=0}^{\infty}z^{-t} Q_{ij}(t)$, which is 
\begin{equation}
\label{eq_Qz}
\widetilde{Q_{ij}}(z)=\left(\mathbf{p^0_i}-\mathbf{p^\infty}\right)\left(\mathbf{I}-\frac{1}{z}\mathbf{S}\right)^{-1}\mathbf{e_{j}}\\
\end{equation}
where $\mathbf{I}$ is the identity matrix. This expression shows that $\widetilde{Q_{ij}}(z)$ is a rational function of z.

We next consider the physical quantity of interest for random walk, namely the first passage probability $F_{ij}(t)$ defined as the probability of reaching the target $j$ for the first time at time $t$. We can first associated $F_{ij}(t)$ with its cumulative distribution function $G_{ij}(t)$, which is the probability that node $j$ has not been visited until time $t$. From these definitions, we see that $F_{ij}(t)=G_{ij}(t-1)-G_{ij}(t)$. We set $F_{ij}(0)=0$, and $G_{ij}(-1)=G_{ij}(0)=1$. From the definition of $F_{ij}(t)$, we have an important relation to describe the dynamics\cite{PhysRevLett.92.118701},
\begin{equation}
P_{ij}(t)=\delta_{t0}\delta_{ij}+\sum_{t'=0}^{t}P_{jj}(t-t')F_{ij}(t')
\end{equation}
We can then use this dynamic equation for $P_{ij}(t)$ and $F_{ij}(t)$ to obtain the key relation between $\widetilde{G_{ij}}(z)$ and $\widetilde{Q_{ij}}(z)$. Note that $\widetilde{Q_{ij}}(z)$ completely determined by $S_{ij}$ by Eq.(\ref{eq_Qz}).
\begin{equation}
\label{eq_Gz}
\widetilde{G_{ij}}(z)=-\frac{\widetilde{Q_{ij}}(z)-\widetilde{Q_{jj}}(z)-\delta_{ij}}{(1-z^{-1})\widetilde{Q_{jj}}(z)+P_{j}^\infty}
\end{equation}
Note that this is the key relation that connects the dynamics with geometry for random walk on any undirected complex network. We observe from this relation that $\widetilde{G_{ij}}(z)$ is solved once $\widetilde{Q_{ij}}(z)$ is known. We can then use inverse Laplace transform to get the cumulative distribution function $G_{ij}(t)$. The usual quantity of interest such as the mean first passage time can easily be computed by $\left<T_{ij}\right>=\sum_{t=0}^{\infty}G_{ij}(t)=\widetilde{G_{ij}}(1)$. In fact, once $G_{ij}(t)$ is known, any other statistical quantity of the random walk on the network will be computable. This is the summary of the formalism for the dynamical equation of the random walk process. 

\section{III. THE INVERSE TRANSFORM OF CUMULATIVE DISTRIBUTION FUNCTION} 
To relate  $\widetilde{G_{ij}}(z)$   to the topological properties of the complex network, we need another equation to compute $\widetilde{Q_{ij}}(z)$, which is a rational function of z and it diverges at all the eigenvalues of $\mathbf{S}$ , except from the eigenvalue $1$. Because $Q_{ij}(t)$ is expected to decrease exponentially, $\widetilde{G_{ij}}(z)$ converges in some region inside $|z|<1$ away from the origin. The inverse Laplace transform of $\widetilde{G_{ij}}(z)$ is,
\begin{equation}
G_{ij}(t)=\frac{1}{2\pi i}\oint\widetilde{G_{ij}}(z)z^{t-1}\mathrm{d}z=\sum_{z_n\in\text{ROC}} Res[\widetilde{G_{ij}}(z_n)z_n^{t-1}]
\end{equation}
The path of integration enclosed the whole region of convergence (ROC). In Appendix A, we find that all poles $z_n$ of $\widetilde{G_{ij}}(z)z^{t-1}$ occurs when the denominator of $\widetilde{G_{ij}}(z)$ is 0, 
\begin{equation}
\label{eq_pole}
(z_n^{-1}-1)\widetilde{Q_{jj}}(z_n)=P_{j}^\infty
\end{equation} 
We call this the pole equation, which can be simplified as a polynomial equation of order $N$, with the corresponding residue having a simple relation to $t$, see Appendix A,
\begin{equation}
\label{eq_residue}
Res[\widetilde{G_{ij}}(z_n)z_n^{t-1}]=A_{z_n}z_n^{t-1}
\end{equation}
where,
\begin{equation}
\label{eq_coeff}
A_{z_n}=-\frac{\widetilde{Q_{ij}}(z_n)-\widetilde{Q_{jj}}(z_n)-\delta_{ij}}{z_n^{-2}\widetilde{Q_{jj}}(z_n)+
(1-z_n^{1})\widetilde{Q_{jj}}'(z_n)}\\
\end{equation}
Note that $\widetilde{Q_{ij}}'(z)$ is the derivative of $\widetilde{Q_{ij}}(z)$. By this equation, we obtain an expansion form of first-passage probability,
\begin{equation}
\label{eq_Gt}
G_{ij}(t)=\sum_{z_n\in\text{ROC}}A_{z_n}z_n^{t-1}
\end{equation}
When $t=0$, the coefficients sum up to $1$, $G_{ij}(0)=\sum_{z_n}A_{z_n}z_n^{-1}=1$. The first-passage cumulative probability distribution function $G_{ij}(t)$ can be represented as finite sum of exponential functions $A_{z_n}z_n^{t-1}$  at poles ${z_n}$. From this observation, we see that an analysis of the poles of $\widetilde{G_{ij}}(z)$ is important for the understanding of the first passage process on networks.

\begin{figure*}[t]
    \centering
    \scriptsize
    \subfigure{
        \begin{minipage}[b]{0.47\linewidth}
            \includegraphics[width=0.95\linewidth]{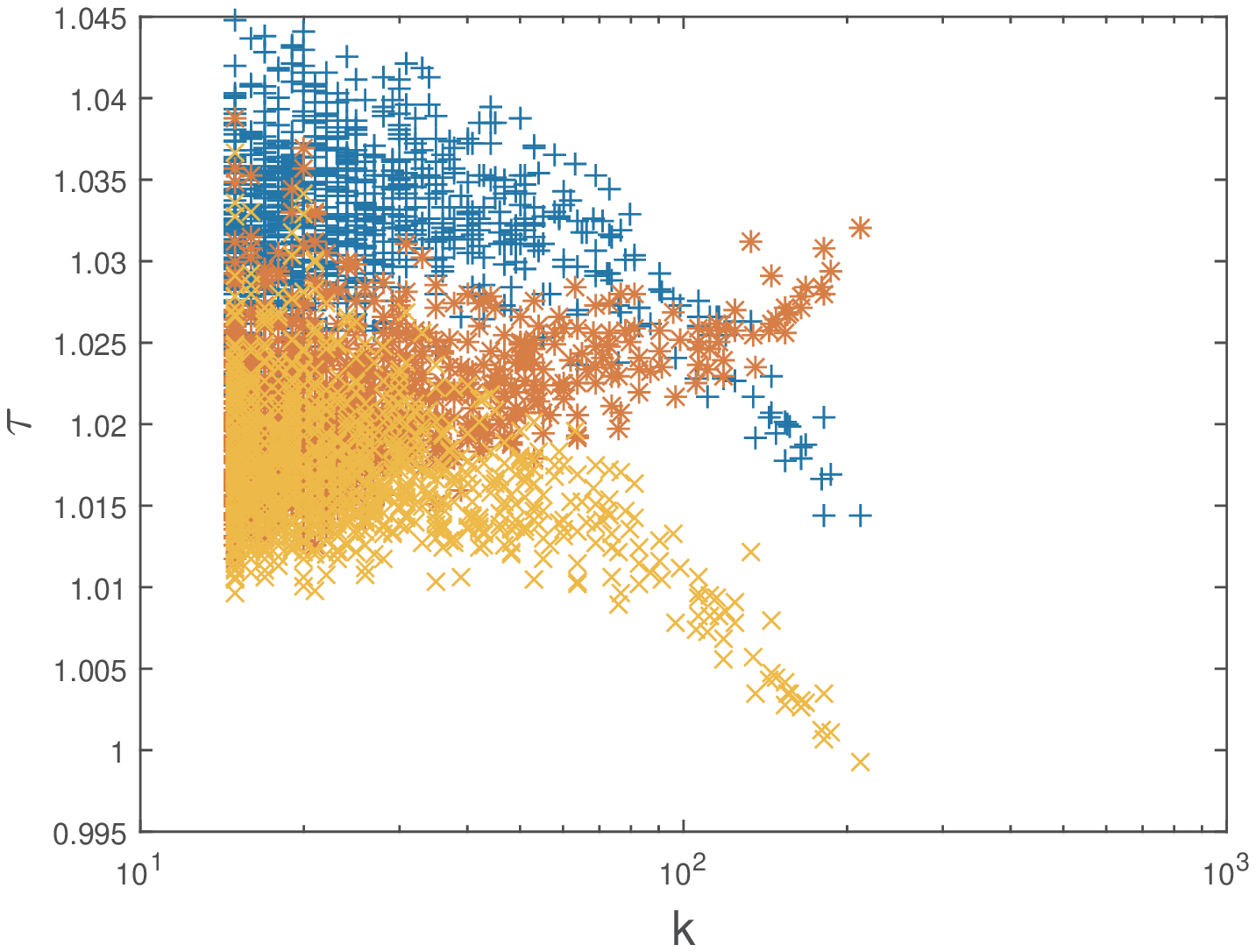} (a)\\
            \includegraphics[width=0.95\linewidth]{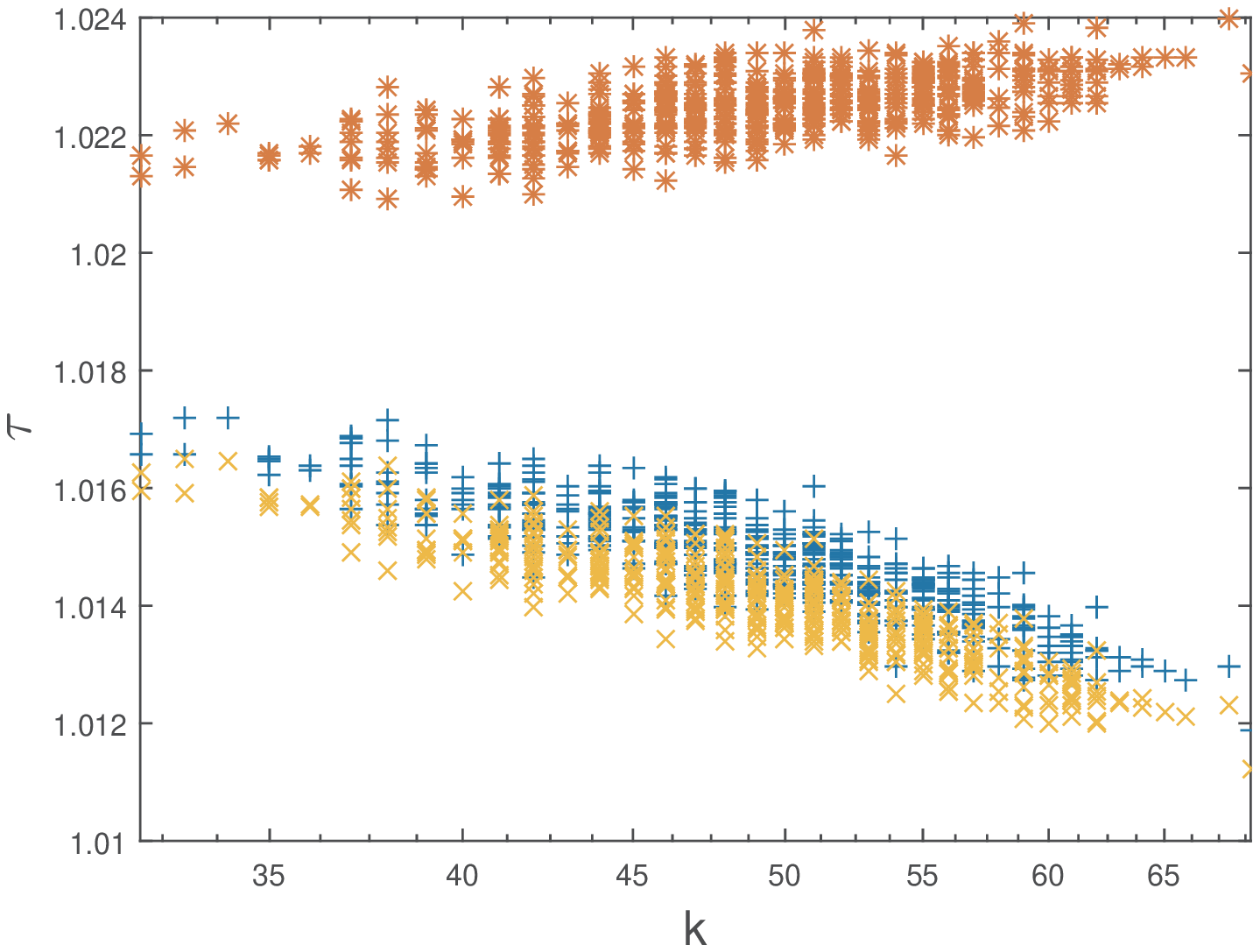} (c)
        \end{minipage}
    }
    \hspace{0.01\linewidth}
    \subfigure{
        \begin{minipage}[b]{0.47\linewidth}
            \includegraphics[width=0.95\linewidth]{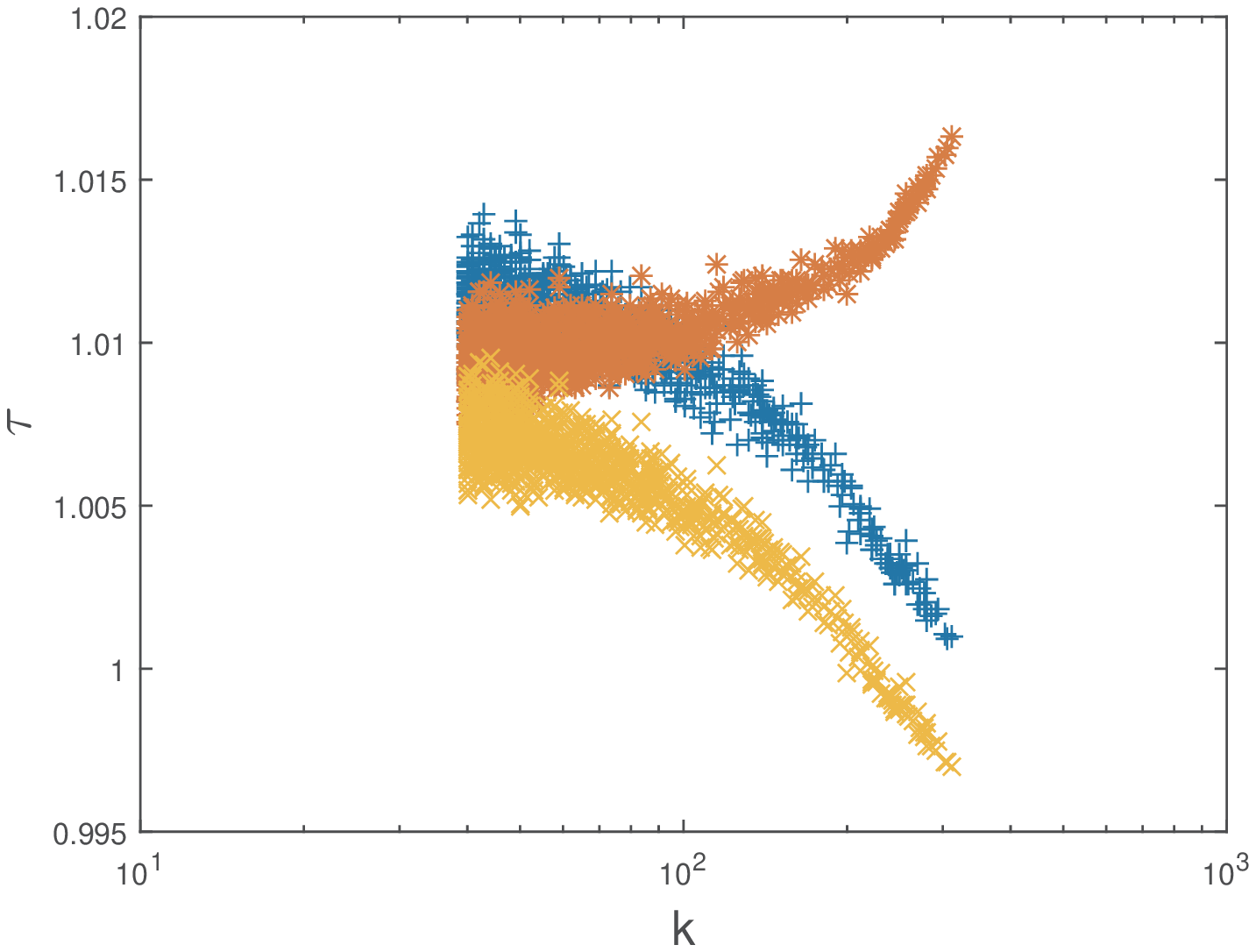} (b)\\
            \includegraphics[width=0.95\linewidth]{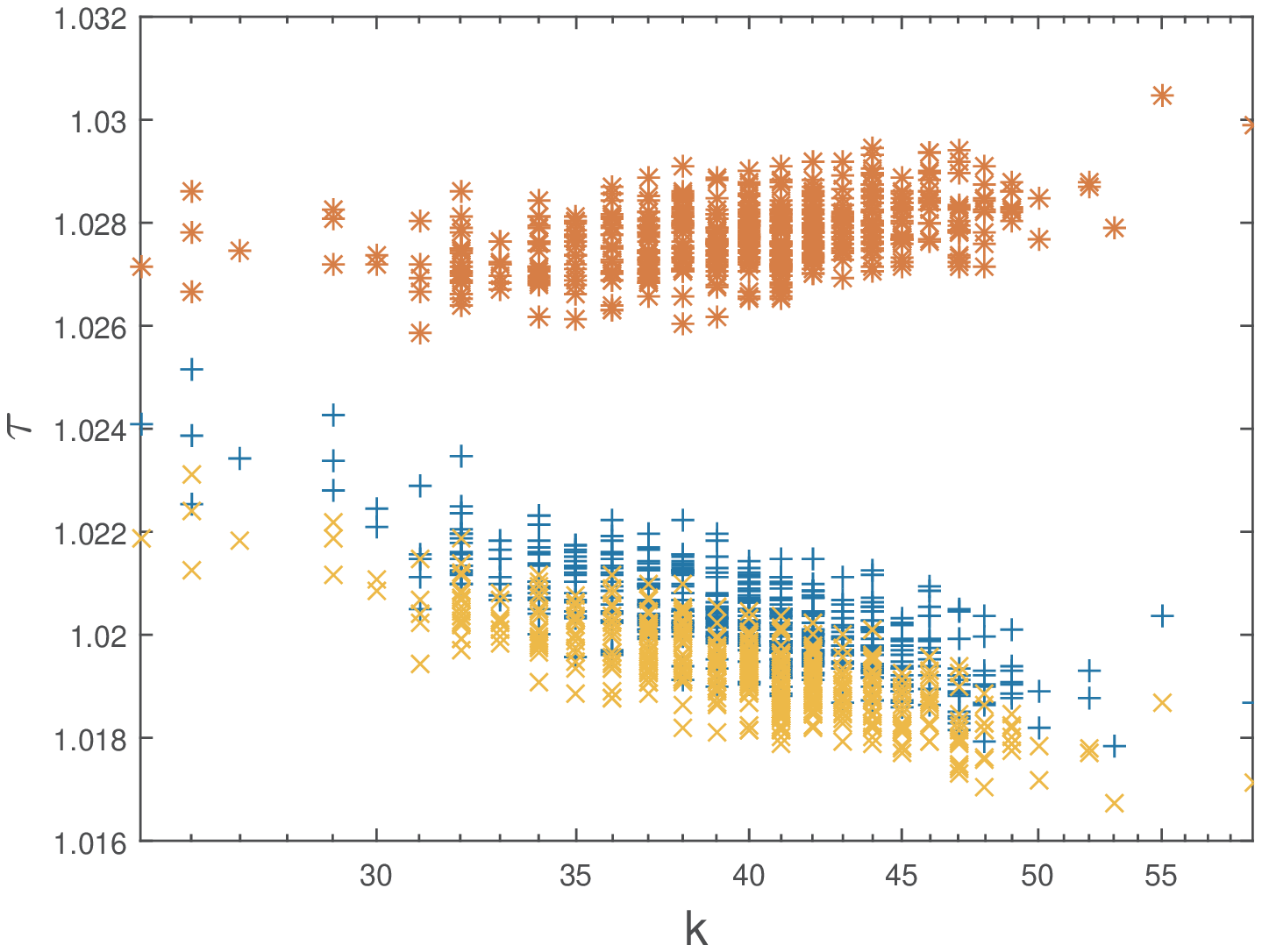} (d)
        \end{minipage}
    }
    \caption{\label{fig_1}Characteristic relaxation time $\tau$ v.s. degree $k$ for all nodes in a network. Blue points (filled circle) for exact simulation results, red (dotted line)  for the simple mean-field approximation, yellow (line) for the improved mean-field approximation. (a): BA network $N=1000$, $m_0=m=15$; (b): BA network $N=1000$, $m_0=m=40$; (c): ER network $N=500$, $\beta=0.1$; (d): WS network $N=500$, $k=20$, $\beta=0.7$}
\end{figure*}

\begin{figure*}[t]
    \centering
    \scriptsize
    \subfigure{
        \begin{minipage}[b]{0.47\linewidth}
            \includegraphics[width=0.95\linewidth]{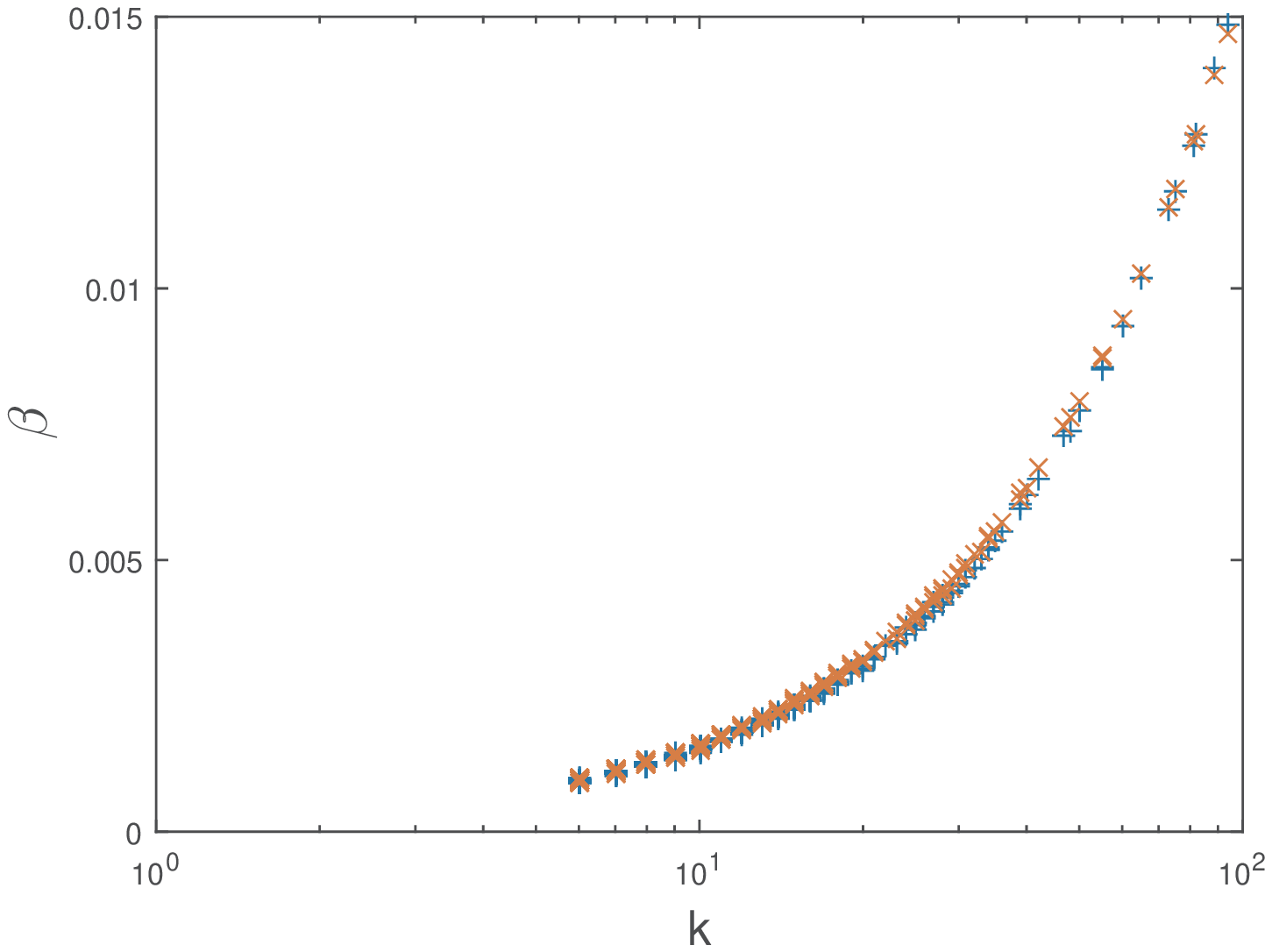} (a)\\
            \includegraphics[width=0.95\linewidth]{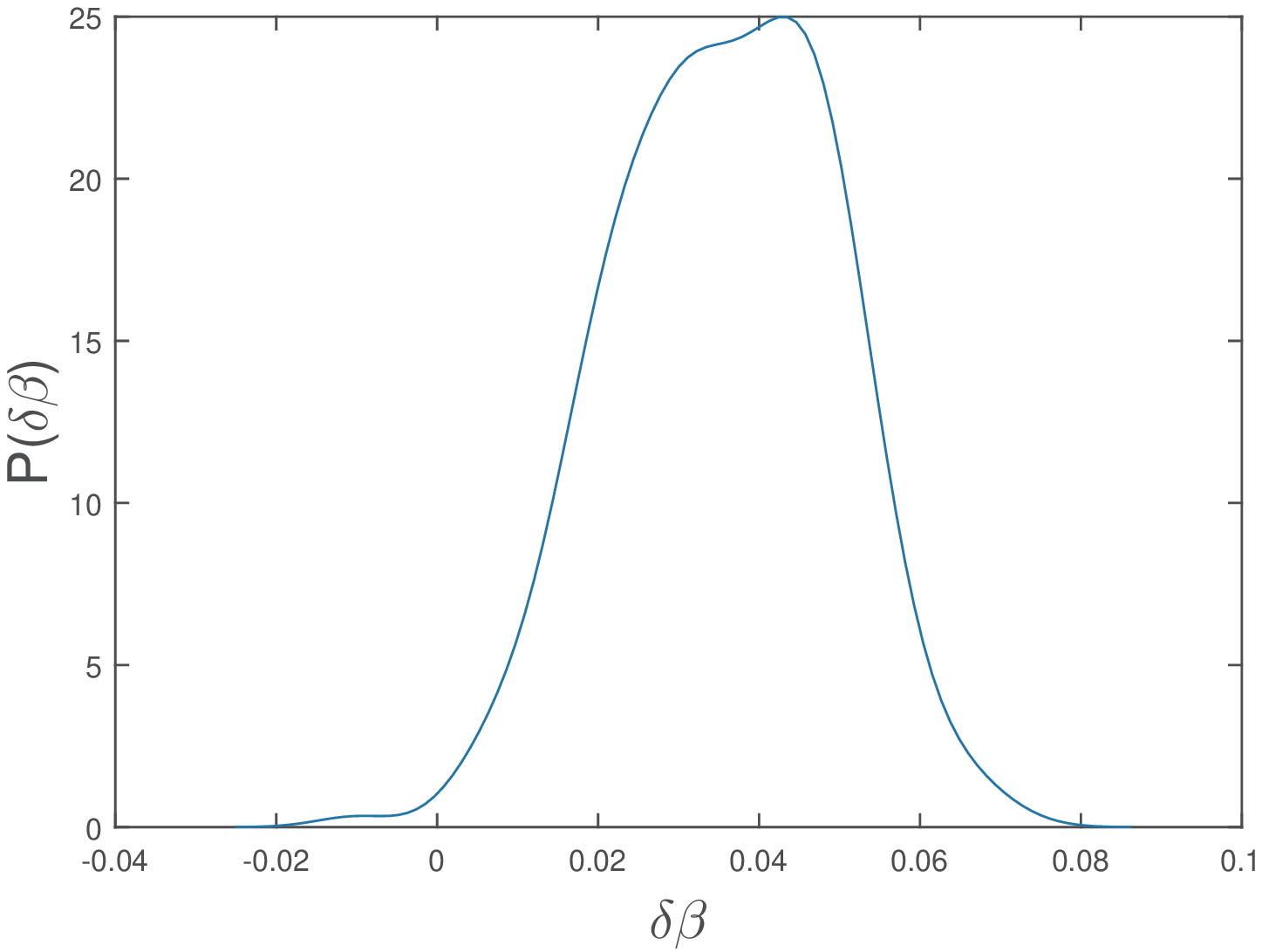} (c)
        \end{minipage}
    }
    \hspace{0.01\linewidth}
    \subfigure{
        \begin{minipage}[b]{0.47\linewidth}
            \includegraphics[width=0.95\linewidth]{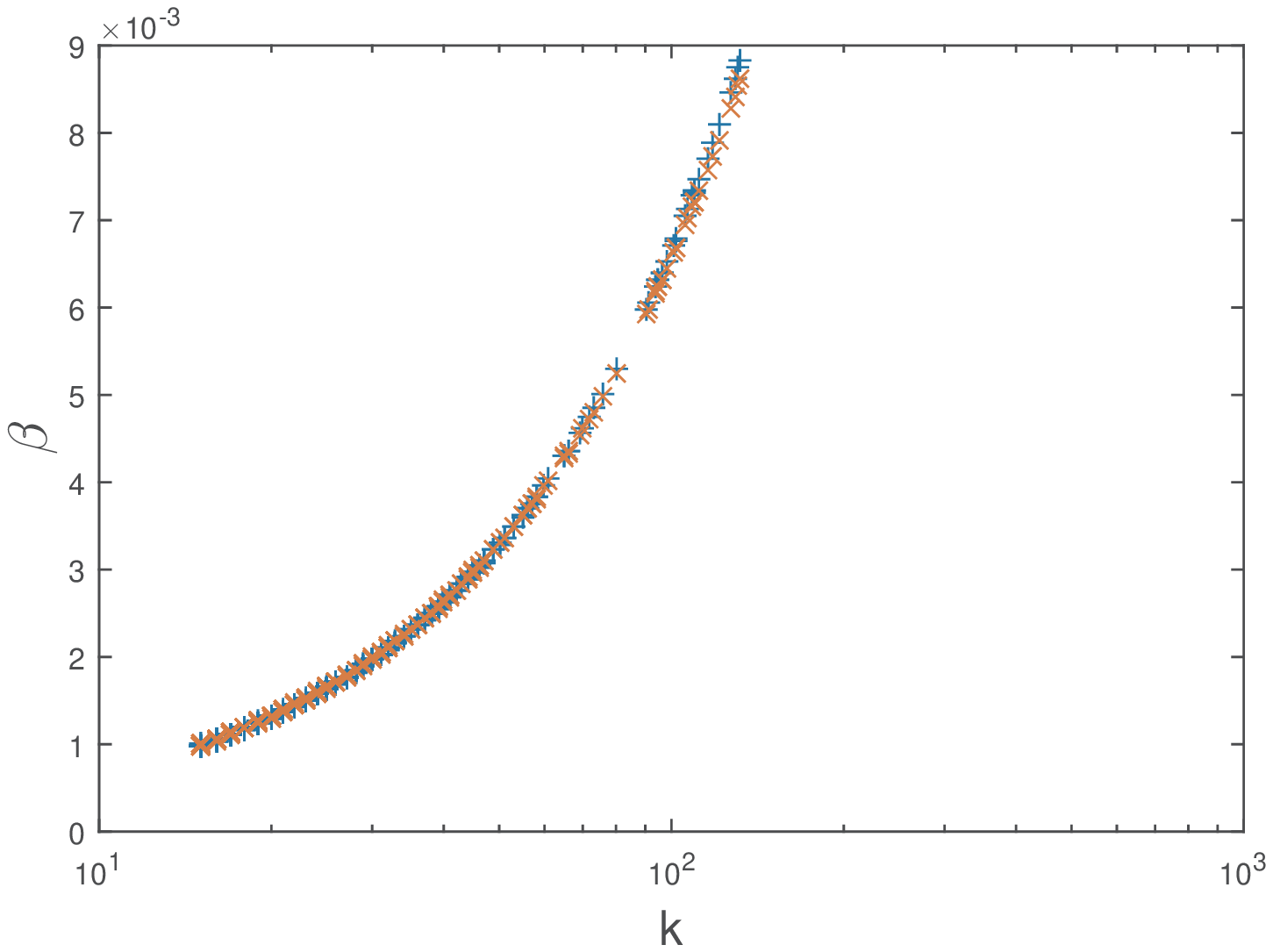} (b)\\
            \includegraphics[width=0.95\linewidth]{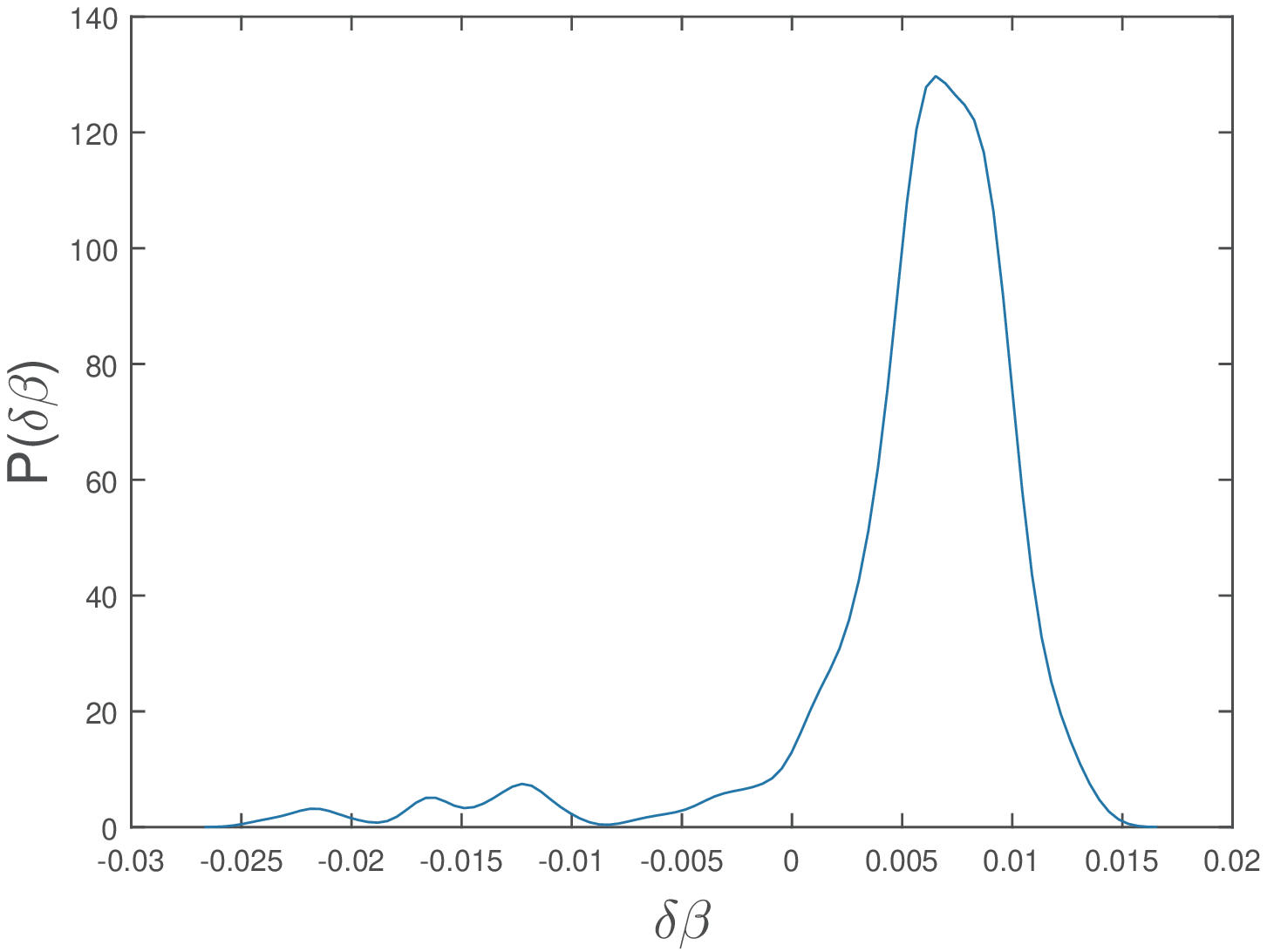} (d)
        \end{minipage}
    }
    \caption{\label{fig_2} (a) and (b): decay rate $\beta$ versus degree $k$ of all nodes in a network. Blue points (filled circle) for exact simulation results, red (dotted line)  for the improved mean-field approximation. (c) and (d): the probability distribution of relative error $\delta\beta$ of decay rate predicted by the improved mean-field approximation in (a) and (b) respectively.
        (a) and (c): BA network $N=500$ and $m=m_0=6$. (b) and (d): BA network $N=500$ and $m=m_0=15$.}
\end{figure*}

\section{V. THE MAIN POLE AND THE ASYMPTOTIC BEHAVIOR OF FIRST-PASSAGE TIME DISTRIBUTION}
We see from the expression of $G_{ij}(t)$ in terms of the summation of terms  $A_{z_n}z_n^{t-1}$ that at a long time, the pole with the largest magnitude $|z_n|$ will dominate the asymptotic behavior of the random walk. We also note that in the equation of poles, Eq.(\ref{eq_pole}), the value of the pole only depends on the target node $j$. This implies that the spectrum is completely determined by the target node $j$ and the topological structure of the local region around it. The relevance of initial node $i$ only appears through the coefficients $A_{z_n}$. We therefore first discuss the main pole, namely the one that determines the asymptotic behavior of the random walk. 

An important feature of the first-passage process for random walk on network is an exponentially decaying asymptotic behavior for $t\to\infty$ \cite{0295-5075-90-4-40005}, with decay rate obtainable by asymptotic analysis with the form, $G_{ij}(t)\approx\mathrm{e}^{-\beta_j t}$, where $\beta_j\lesssim k_i/ (N\left<k\right>)=P_{j}^\infty$. 
This implies that there should be a positive real pole $z_0$ slightly less than $1$, such that, 
$z_0=\mathrm{e}^{-\beta_j}\lesssim 1$ and $ A_{z_0}\lesssim 1$.
Because the asymptotic behavior is dominated by $z_0$, all the other poles $z_n$ satisfies, $|z_n|<<z_0$ for $n\neq0$.

We can verify the above identification of the main pole $z_0$, by solving the pole equation Eq.(\ref{eq_pole}),  
We first note that for a vast network, the right-hand side of the pole equation, $P_{j}^\infty$, is minimal.  
In order for the pole equation to be valid, we must have the left-hand side of the pole equation, 
$(z_n^{-1}-1)\widetilde{Q_{jj}}(z_n)$,  also very small. 
In this expression, $(z_n^{-1}-1)$ can be small when $ z_n \to 1$, 
but  $\widetilde{Q_{jj}}(1)$ needs not be small in general, as $\widetilde{Q_{jj}}(z)$ converges and $\widetilde{Q_{jj}}(1)\gg\beta_j \lesssim k_i/(N\left<k\right>)=P_{j}^\infty$. 
Therefore, we can say that there should be a pole $z_0$ with magnitude slightly less than 1. By expanding all terms to the first order of $(z-1)$, we get  
\begin{equation}
z_0\approx 1+\frac{P_{j}^\infty}{\frac{\mathrm{d}}{\mathrm{d}z}\left[(\frac{1}{z}-1)\widetilde{Q_{jj}}(z)\right]\Big|_{z=1}}
\approx\mathrm{e}^{-\frac{P_{j}^\infty}{\widetilde{Q_{jj}}(1)}}\\
\end{equation}
where $\widetilde{Q_{jj}}(1)=\sum_{t=0}^{\infty}(P_{jj}(t)-P_{j}^\infty)$ is identified as the characteristic relaxation time $\tau_j$. 
The decay rate for the asymptotic behavior of the random walk will thus be 
\begin{equation}
\label{eq_decayRate}
\beta_j\approx\frac{P_{j}^\infty}{\tau_j}
\end{equation}
As we can see later, this formula gives a very accurate approximation of the decay rate $\beta_j$. 
The next task is to compute the relaxation time $\tau_j =\widetilde{Q_{jj}}(1)$.

\section{VI. MEAN-FIELD APPROXIMATION}
We first discuss a simple version of the mean-field approximation. Since the characteristic relaxation time $\tau_{j}$ is a function of node $j$, we may estimate $\tau_{j}$ by the local topological structure around the node $j$. We may only consider the topological information within the nearest neighbors of the node $j$ and divide the whole network into three disjoint regions: region $(0)$ is just the node $j$ itself, region $(1)$ consists of the $k_j$ nearest neighbors of node $j$, and region $(\infty)$ consists of the remaining nodes in the network. In this simple mean-field approximation, we treat the state $(\infty)$ as an absorbing state. For a random walker, we need two independent parameters to represent all the transition probabilities describing his move between these three regions. The first parameter $p_1$ concerns a random walker going from region $(1)$ to region $(0)$. The second parameter $p_2$ describes the probability that the random walker remains in the nearest neighbor region $(1)$. The stochastic process for this random walker is described by the transition matrix as follows, 
\[
\bordermatrix{
& 0& 1&\infty \cr
0& 0& 1&0 \cr
1& p_1&p_2&1-p_1-p_2 \cr
\infty &0&0&1 \cr}
\]
We see that the probability of moving from region $(1)$ to $(\infty)$ must be $(1-p_1-p_2)$, since the sum of transition probabilities from the state $(1)$ to all the other states is $1$.

\begin{figure}[h]
    \centering
    \scriptsize
    \includegraphics[width=0.35\textwidth]{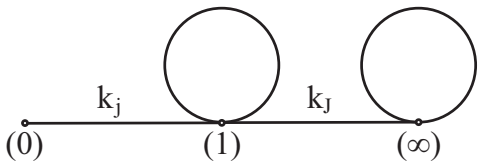}
    \caption{\label{fig_0}Schematic diagram of the network structure under mean-field approximation. Note that each edge in the diagram actually represent a set of several edges. The number of edges connecting region $(0)$ and $(1)$ is $k_j$. The number of edges connecting region $(1)$ and region $(\infty)$ is the degree of the giant node $J$ formed by region $(0)$ and $(1)$ together, which is denoted as $k_J$. Note that there are loops at the state $(1)$ and $(\infty)$.}
\end{figure}

Now we want to estimate the two parameters $p_1$ and $p_2$ by the local topological information. The structure of the reduced network in mean-filed approximation is schematically shown in FIG.(\ref{fig_0}). We can write out the transition probability $p_1$ and $p_2$ directly from this reduced network. Note that we shall think there are infinite number of loops at state $(\infty)$ since we regard it as an absorbing state. The degree of the abstract node $(1)$ is the sum of degrees of all nodes in region $(1)$, which is $\sum_{\nu\in\text{Nei}(j)}k_\nu=k_j\left<k_\nu\right>$. Note that $\left<k_\nu\right>$ denotes the average degree of all nearest neighbors of node $j$. And the number of edges connecting abstract node $(1)$ and node $(\infty)$ is the degree of the giant node formed by region $(0)$ and $(1)$,
\begin{equation}
k_J=k_j\left<k_\nu\right>-k_j-c_jk_j(k_j-1)
\end{equation}
where $c_j$ is the clustering coefficient of node $j$. We shall make an assumption that in this reduced network, the random walker still move through any edge with equal probabilities, as an approximation. Then the parameters $p_1=k_j/\sum_{\nu\in\text{Nei}(j)}k_\nu=1/\left<k_\nu\right>$ and $p_2=1-p_1-k_J/\sum_{\nu\in\text{Nei}(j)}k_\nu=c_j(k_j-1)/\left<k_\nu\right>$.

Because the $\infty$ state is an absorbing state, the stochastic matrix of simple mean-field approximation can be future reduced to a $2\times2$ matrix. 
We obtain the formula of $P_{jj}(t)$ as,
\begin{equation}
\begin{pmatrix}
P_{jj}(t),&\sum_{\nu\in\text{Nei}(j)}P_{j\nu}(t)\\
\end{pmatrix}=
\begin{pmatrix}
1&0\\
\end{pmatrix}
\begin{pmatrix}
0&1\\
p_1&p_2\\
\end{pmatrix}^t
\end{equation}
By Eq.(\ref{eq_Qz}), we can readily obtain,
\begin{equation}
\widetilde{Q_{jj}}(z)=\begin{pmatrix}
1&0\\
\end{pmatrix}\begin{pmatrix}
1&-\frac{1}{z}\\
-\frac{p_1}{z}&1-\frac{p_2}{z}\\
\end{pmatrix}^{-1}\begin{pmatrix}
1\\0\\
\end{pmatrix}
\end{equation}
And one can derive the approximate formula of the characteristic relaxation time $\tau_j=\widetilde{Q_{jj}}(1)$,
\begin{equation}
\label{eq_relaxationTime}
\tau_j=\frac{1-p_2}{1-p_1-p_2}
\end{equation}
Now, with Eq.(\ref{eq_decayRate}) and Eq.(\ref{eq_relaxationTime}), we obtain an approximate formula of the decay rate $\beta_j$. Simulations have verified these analytical results. 

The simple mean-field theory can be improved by introducing a non-zero returning probability $p_3$ from state $(\infty)$ to state $(1)$. Thus the improved mean-field approximation uses the following transition matrix, 
\[
\bordermatrix{
    & 0& 1&\infty \cr
    0& 0& 1&0 \cr
    1& p_1&p_2&1-p_1-p_2 \cr
    \infty &0&p_3&1-p_3 \cr}
\]
The critical change in this improved mean-field approximation is that the transition matrix now completely describes the reduce network in FIG.(\ref{fig_0}), which is self-consistent since it describes a legal undirected network. However, we have to use some global information of the original network, namely the size $N$ and the average degree $\left<k\right>$, to calculate $p_3$. The degree of abstract node $(\infty)$ is the sum of degrees of all nodes in region $(\infty)$, which is $(N\left<k\right>-k_j-k_j\left<k_\nu\right>)$. Then we have $p_3=k_J/(N\left<k\right>-k_j-k_j\left<k_\nu\right>)$.

By Eq.(\ref{eq_Qz}), the corresponding solution for $\widetilde{Q_{jj}}(z)$ is,
\begin{equation}
\begin{split}
&\widetilde{Q_{jj}}(z)=\\
&\begin{pmatrix}
1-P_0^\infty , &-P_1^\infty, &-P_\infty^\infty\\
\end{pmatrix}\begin{pmatrix}
1&-\frac{1}{z}&0\\
-\frac{p_1}{z}&1-\frac{p_2}{z}&-\frac{1-p_1-p_2}{z}\\
0&-\frac{p_3}{z}&1-\frac{1-p_3}{z}\\
\end{pmatrix}^{-1}\begin{pmatrix}
1\\0\\0\\
\end{pmatrix}
\end{split}
\end{equation}
Similarly one can calculate the characteristic relaxation time $\tau_j=\widetilde{Q_{jj}}(1)$ as,
\begin{equation}
\tau_j=\frac{p_1(-1+p_2-2p_3)+(1-p_2+p_3)^2}{(1-p_2+p_1(-1+p_3)+p_3)^2}
\end{equation}

More interestingly, the improved mean-field approximation describes a legal network as in FIG.(\ref{fig_0}). If we use the approximate $\widetilde{Q_{jj}}(z)$ function in the equation of poles Eq.(\ref{eq_pole}) at node $j$, we are guaranteed to find an approximate main pole $z_0$. The equation of poles is a quadratic equation of $z$,
\begin{equation}
\begin{split}
&\frac{p_1+p_2-p_3-z}{p_1 (p_3-1)+z (-p_2+p_3+z)}\\
&+\frac{p_1+p_2-p_3-1}{-(p_1+1)
p_3+p_1+p_2-1}+P_j^\infty=0\\
\end{split}
\end{equation}
Numerical calculations have verified that there is a solution $z_0\lesssim 1$ corresponding to the main pole. In next section we show that this improved mean-field theory provides excellent agreement with simulation.

\begin{figure*}[t]
    \centering
    \scriptsize
    \subfigure{
        \begin{minipage}[b]{0.47\linewidth}
            \includegraphics[width=0.95\linewidth]{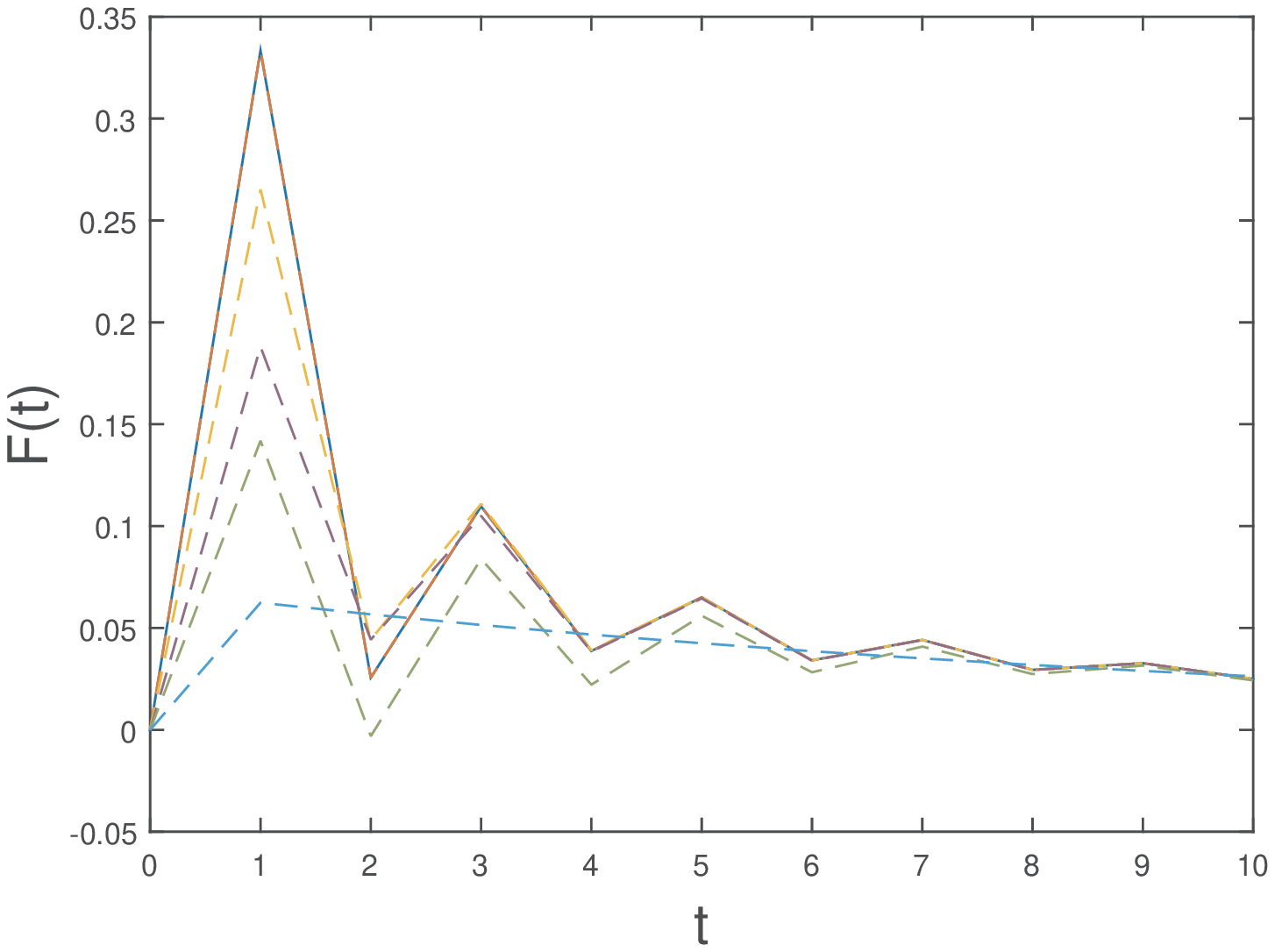} (a)\\
        \end{minipage}
    }
    \hspace{0.01\linewidth}
    \subfigure{
        \begin{minipage}[b]{0.47\linewidth}
            \includegraphics[width=0.95\linewidth]{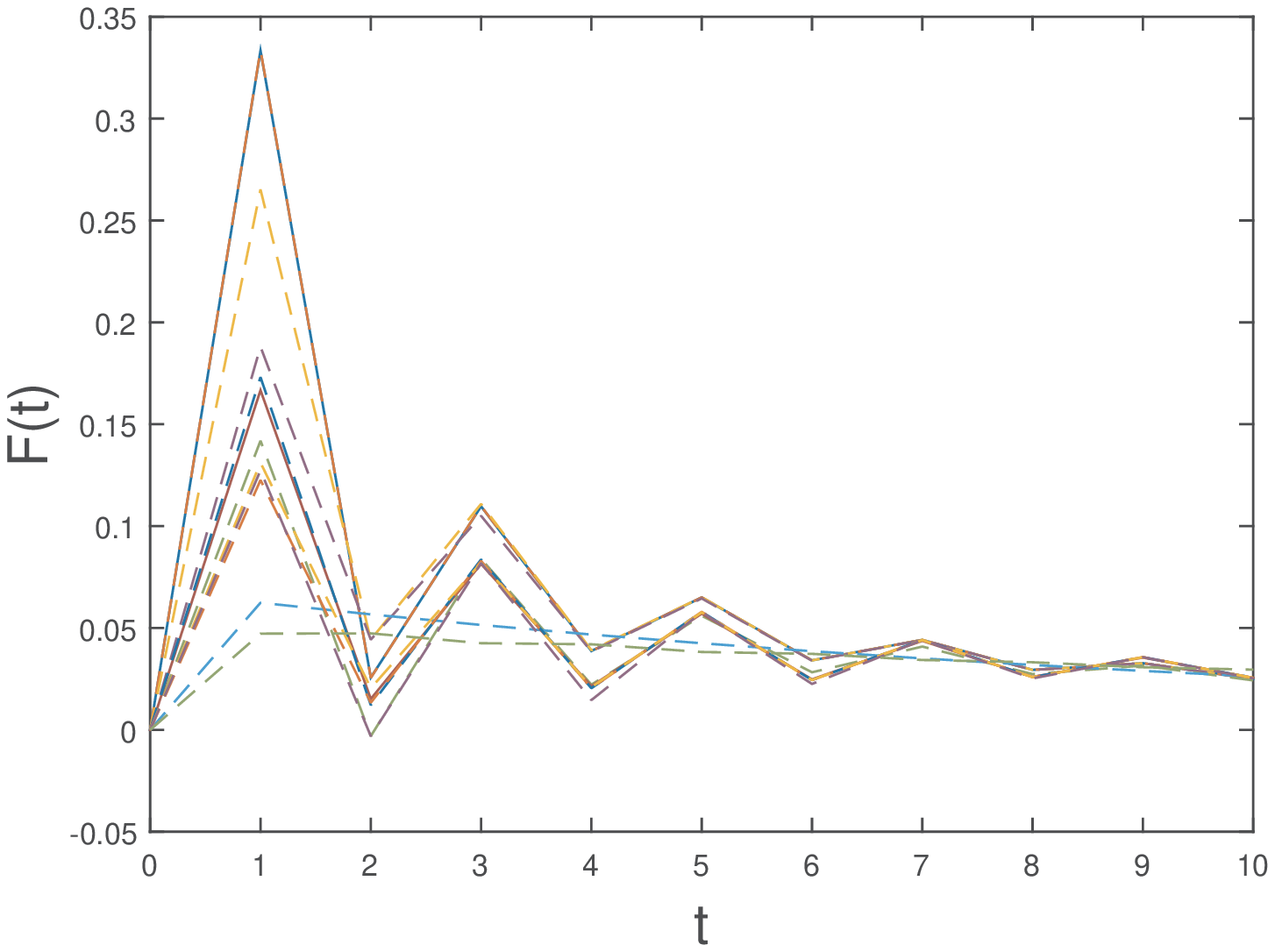} (b)\\
        \end{minipage}
    }
    \caption{\label{fig_3} First-passage time probability distribution $F(t)$ v.s. time $t$ for a random node in a network. The exact first passage time distribution function (black line), and the predictions by the expansion formula (dashed lines) when omitting the poles whose absolute value is less than $0.2$, $0.4$, $0.6$, $0.8$, respectively. (a): BA network $N=30$, $m_0=m=2$ ; (b): BA network $N=100$, $m_0=m=2$.}
\end{figure*}

\begin{figure*}[t]
    \centering
    \scriptsize
    \subfigure{
        \begin{minipage}[b]{0.47\linewidth}
            \includegraphics[width=0.95\linewidth]{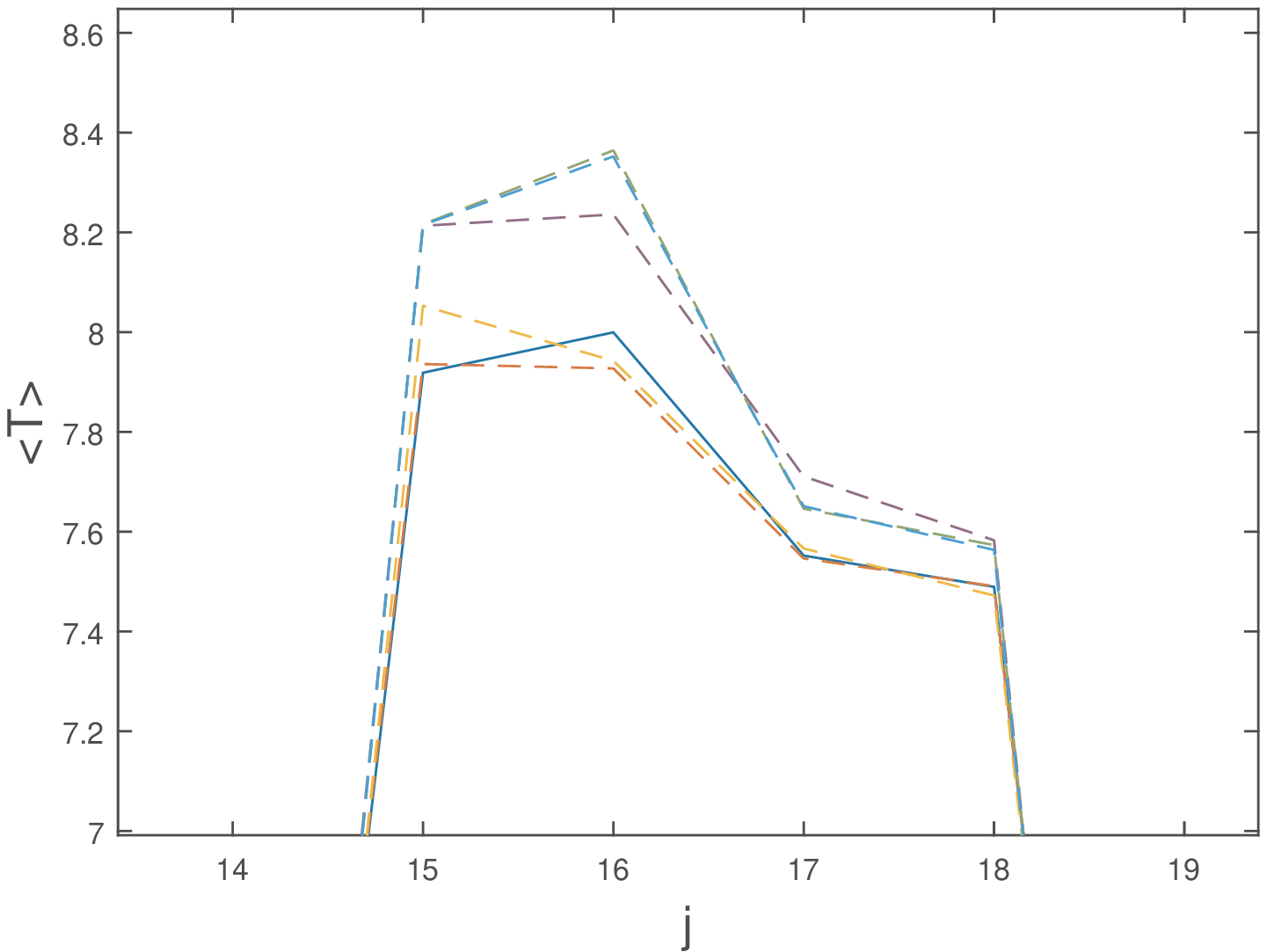} (a)\\
            \includegraphics[width=0.95\linewidth]{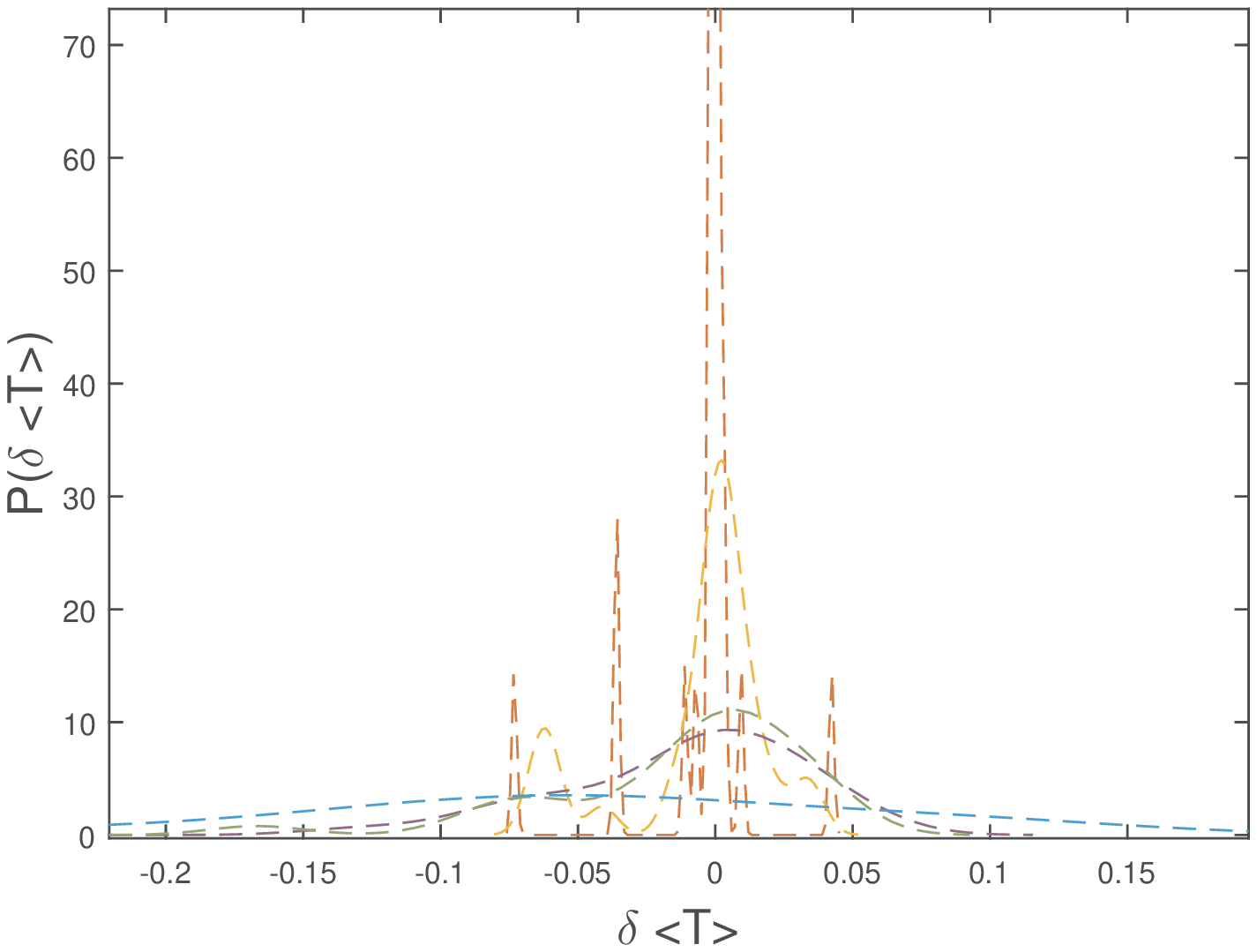} (c)
        \end{minipage}
    }
    \hspace{0.01\linewidth}
    \subfigure{
        \begin{minipage}[b]{0.47\linewidth}
            \includegraphics[width=0.95\linewidth]{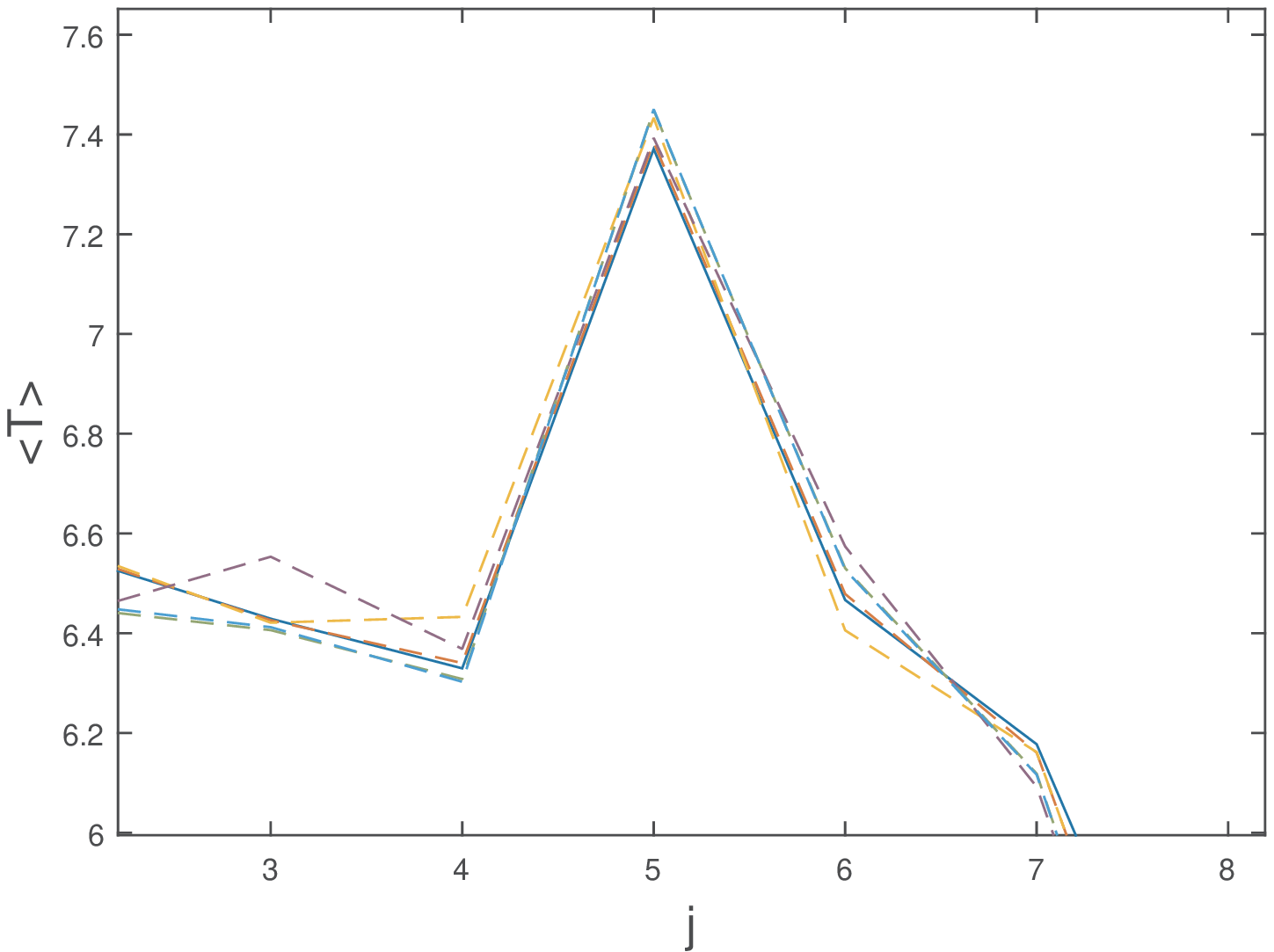} (b)\\
            \includegraphics[width=0.95\linewidth]{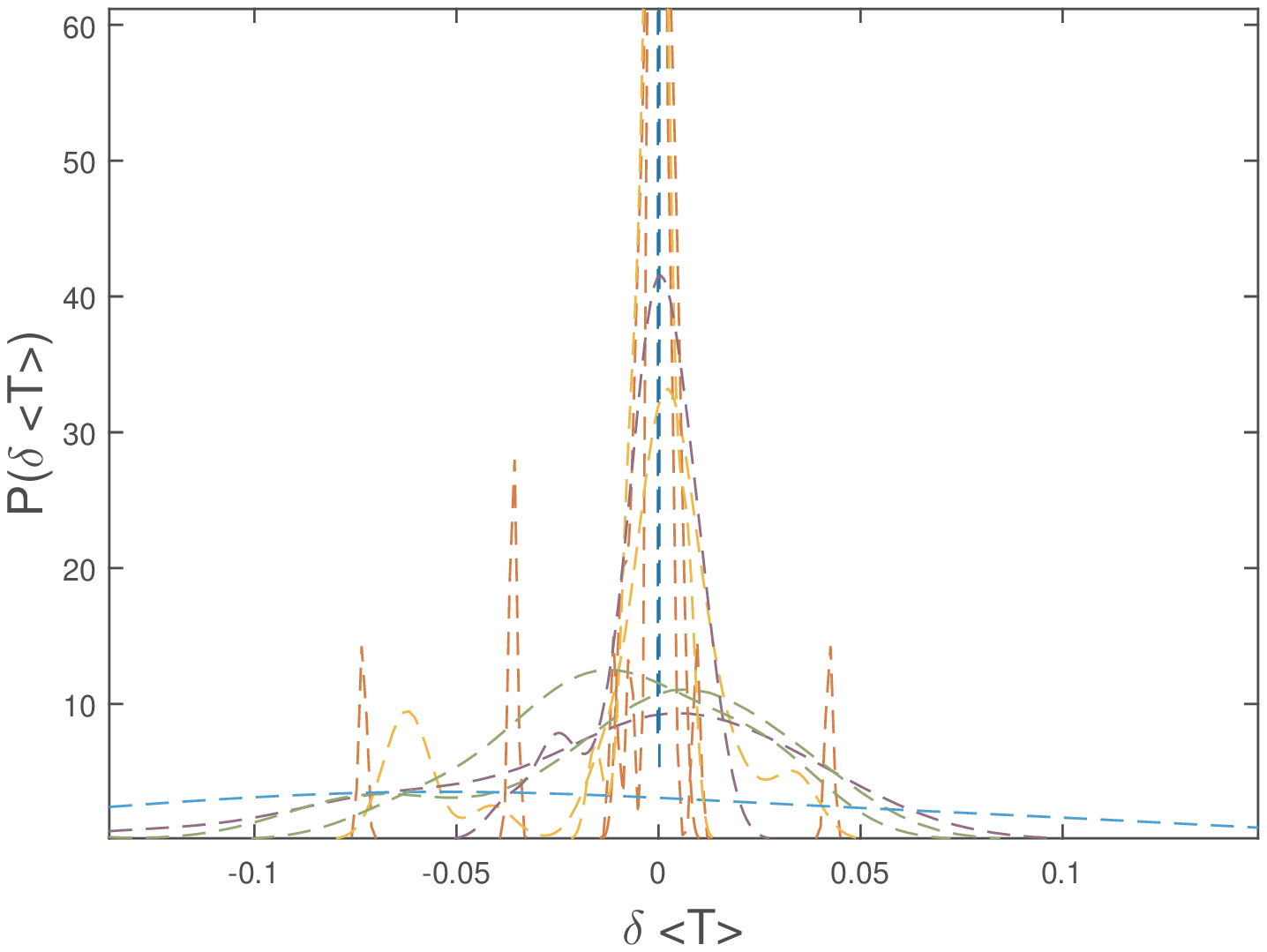} (d)
        \end{minipage}
    }
    \caption{\label{fig_4}(a) and (b): part of the graph of mean first passage time $\left<T\right>$ v.s. NO. of the target node $j$. The black line represents the exact value and the dashed lines are the prediction by expansion formula when omitting the poles whose absolute value is less than $0.2$, $0.4$, $0.6$, $0.8$, respectively. (c) and (d): the probability distribution of relative error $\delta\left<T\right>$ predicted by expansion formula in (a) and (b) respectively. The dashed lines are in one-to-one correspondence to those in (a) and (b) according to their colors. (a): BA network $N=30$, $m_0=m=2$; (b): BA network $N=30$, $m_0=m=2$.}
\end{figure*}

\begin{figure*}[t]
    \centering
    \scriptsize
    \subfigure{
        \begin{minipage}[b]{0.47\linewidth}
            \includegraphics[width=0.95\linewidth]{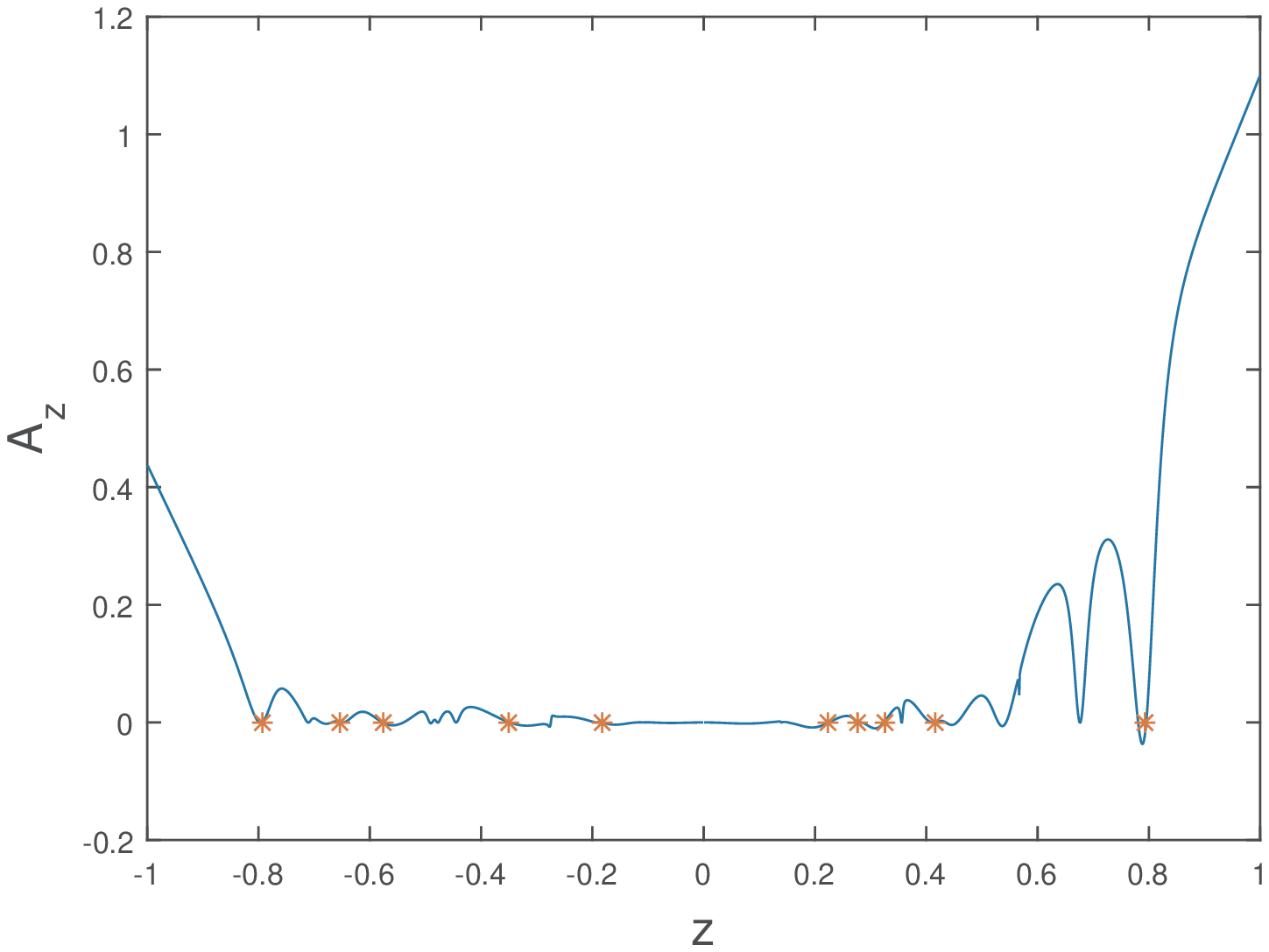} (a)\\
        \end{minipage}
    }
    \hspace{0.01\linewidth}
    \subfigure{
        \begin{minipage}[b]{0.47\linewidth}
            \includegraphics[width=0.95\linewidth]{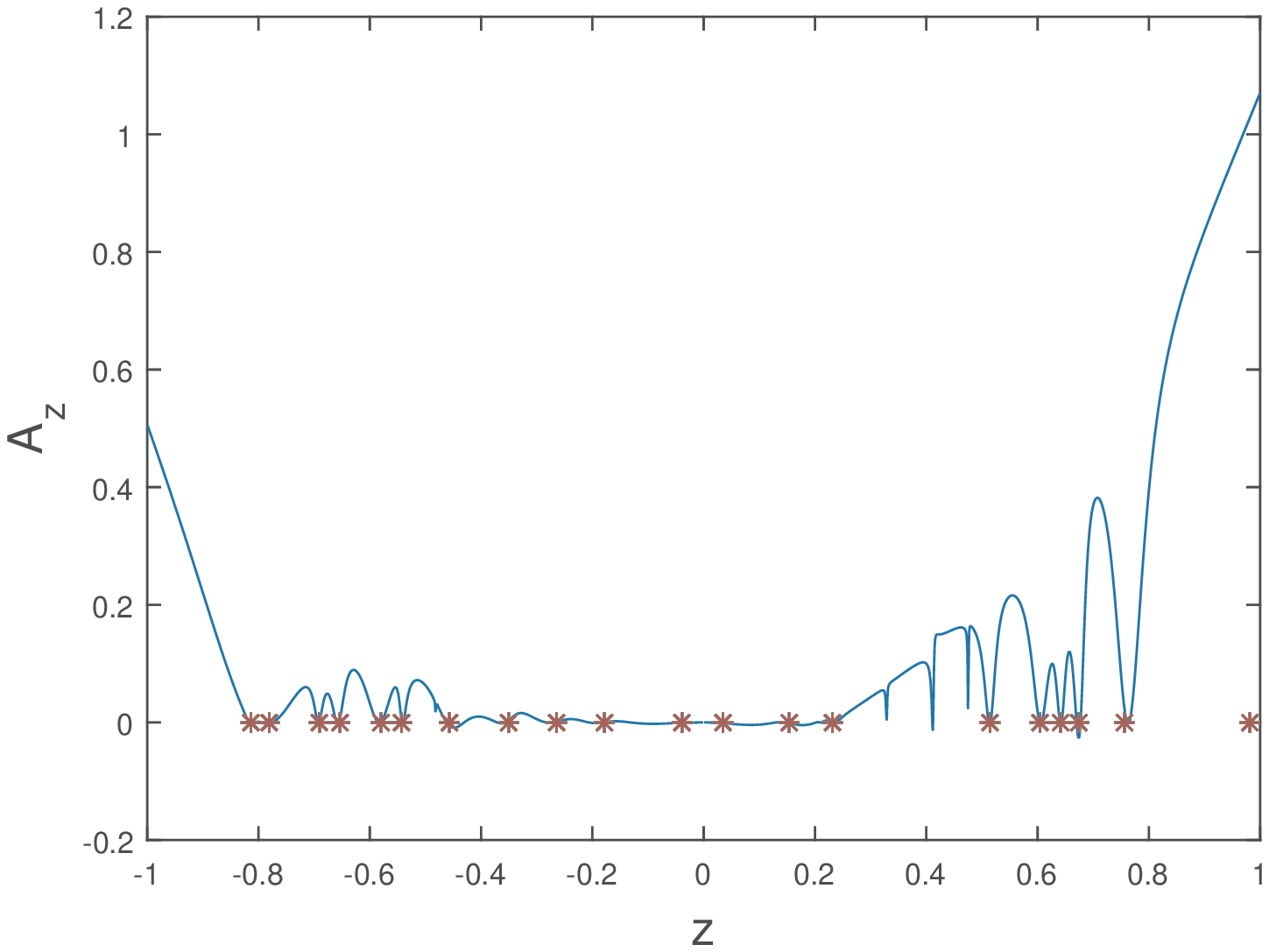} (b)\\
        \end{minipage}
    }
    \caption{\label{fig_5}The coefficient function $A_z$ v.s. z, where zero value corresponds to a pole, in BA network with parameter $N=30$ and $m_0=m=2$. (a): The red stars in graph indicate the singular points of $\widetilde{Q_{jj}}(z)$; (b): The green stars in graph indicate the poles}
\end{figure*}

\section{VIII. COMPARISON WITH NUMERICAL RESULTS}

\subsection{CHARACTERISTIC RELAXATION TIME}
The approximate values of the characteristic relaxation time by two versions of mean-field are compared to the exact results by simulations, see in FIG.(\ref{fig_1}).

We can find that both models work well when the degrees of nodes are relatively small, but for high degree nodes, the improved version is unquestionably better. This effect makes sense because a node with a high degree has a large cluster so that treating the region $\infty$ as the absorbing state is not appropriate.

\subsection{THE DECAY RATE}
The improved mean-field model gives a very accurate prediction of the decay rate, see in FIG.(\ref{fig_2}).

\subsection{FPT DISTRIBUTION AND THE MFPT}
The Expansion form of first passage time distribution function allows us to obtain various approximate formulas of the distribution. It is clear the importance of poles $z_n$ is ordered by their absolute values $|z_n|$. The more poles with small absolute values we take into consideration, the more detail information of the first passage distribution we can get. See FIG.(\ref{fig_3}) for some typical cases.

From the results above we find that the prediction of first passage time distribution better than the asymptotic approximation. This approach should also result in a more accurate prediction of the mean first passage time, as shown in FIG.(\ref{fig_4}).

The pole expansion form gives a more accurate formula of the mean first passage time, which is essential when an accurate prediction of the mean first passage time is needed.

\section{XI. THE COEFFICIENTS IN THE EXPANSION FORM}
We find the coefficient function $A_z$ has some special patter in function graph, as shown in FIG.(\ref{fig_5}),

From the formula of coefficient function $A_z$, Eq.(\ref{eq_coeff}), we know that the coefficient $A_z$ reaches $0$ when $\widetilde{Q_{jj}}(z)$ or $\widetilde{Q_{jj}}'(z)$ is at its maximum or singular point.
The singular points of $\widetilde{Q_{jj}}(z)$ do indicate most of the zero points of $A_{z}$, while the others correspond to the maximum points of $\widetilde{Q_{jj}}'(z)$ between two singular points of $\widetilde{Q_{jj}}(z)$.
So we can conclude that the flat bottom of the coefficient function $A_z$ is because of the distribution of the singular point of $\widetilde{Q_{jj}}(z)$. The range of flat bottom is just the range between the smallest and the most significant singular point. 
We found that except from the main pole $z_0$, all the other poles locate between the zero points of $A_z$. Because the FPT distribution function in z domain $\widetilde{Q_{jj}}(z)$ is monotonic between any two singular points along the real axis, so there will be one or no pole between every two neighbor singular points, i.e. every two zero points of coefficient $A_z$.
This shows that only the main pole can have large coefficient, 
$A_{z_0}>>A_{z_i}$ for $i\neq0$, which validates our prediction, $A_{z_0}\lesssim 1$.

\section{XII. CONCLUSION}
In summary, we studied the first-passage probability at a given time. By the inverse Laplace transform, we could write the probability as the summation of finitely many terms with different time dependence. Among those terms, we identified the one correspond to the asymptotic decaying\cite{0295-5075-90-4-40005} and obtain an accurate formula for the decay rate $\beta$. We developed a mean field approximation method to estimate the characteristic relaxation time by local region network information, and the theoretical results are well supported by numerical simulations of BA, ER, and WS network. By analysis the rest terms in the formula of first-passage probability, it is possible to find some structural parameters of the network which are critical to the first-passage process. 

\appendix*

\section{\label{appendix}APPENDIX: POLES AND RESIDUES OF $\widetilde{G_{ij}}(z)z^{t-1}$ IN THE INVERSE LAPLACE TRANSFORM}
By Eq.(\ref{eq_Gt}), the cumulative distribution function of the first-passage time, $G_{ij}(t)$, is a function of all residues of $\widetilde{G_{ij}}(z)z^{t-1}$ in the region of convergence (ROC). From Eq.(\ref{eq_pole}) and Eq.(\ref{eq_Qz}), we know that all the poles are completely determined by the stochastic matrix $\mathbf{S}$. It is clear that its determinant $\det\left(z\mathbf{I}-\mathbf{S}\right)$ and each element of its matrix of cofactors $\mathbf{C}$ are polynomials of $z$, and the order of $\det\left(z\mathbf{I}-\mathbf{S}\right)$ as a polynomial is exactly $N$.
We rewrite the Eq.(\ref{eq_Qz}) in a new way as, 
\begin{equation}
\label{eq_Qz2}
\widetilde{Q_{ij}}(z)=\frac{z\left(\mathbf{p^0_i}-\mathbf{p^\infty}\right)\mathbf{C}^\intercal\mathbf{e_{j}}}{\det\left(z\mathbf{I}-\mathbf{S}\right)}\\
\end{equation}
Now we can see that $\widetilde{Q_{ij}}(z)$ is a rational function of $z$. And especially, the denominator $\det\left(z\mathbf{I}-\mathbf{S}\right)$ is irrelevant to $i$ and $j$. From Eq.(\ref{eq_Gz}), the formula of $\widetilde{G_{ij}}(z)$, we know that this denominator can be eliminated and makes the numerator in Eq.(\ref{eq_Gz}) be simply a polynomial of $z$. Thus, we conclude that the poles appear only when the denominator of $\widetilde{G_{ij}}(z)$ is zero, which is exactly Eq.(\ref{eq_pole}).
By Eq.(\ref{eq_pole}) and Eq.(\ref{eq_Qz2}), we can rewrite the equation of poles as,
\begin{equation}
(1-z)\left(\mathbf{p^0_i}-\mathbf{p^\infty}\right)\mathbf{C}^\intercal\mathbf{e_{j}}=P_{j}^\infty\det\left(z\mathbf{I}-\mathbf{S}\right)
\end{equation}
Usually this polynomial equation cannot be reduced further as it is of order $N$. This implies that there are at most $N$ distinct poles.
Now suppose at $z_n$, there is a pole of order $m$ of $\widetilde{G_{ij}}(z)z^{t-1}$, then the residues at $z_n$ is,
\begin{equation}
\begin{split}
&Res[\widetilde{G_{ij}}(z_n)z_n^{t-1}]\\&=\frac{1}{(m-1)!}\lim_{z\to z_n}\frac{\mathrm{d}^{m-1}}{\mathrm{d}z^{m-1}}[(z-z_n)^m\widetilde{G_{ij}}(z)z^{t-1}]
\end{split}
\end{equation}
Since $\widetilde{Q_{ij}}(z)$ is usually smooth when it converges, we assume that at the most of times the roots of Eq.(\ref{eq_pole}) have multiplicity one. Thus most of the poles are single. This time we have,
\begin{equation}
\begin{split}
Res[\widetilde{G_{ij}}(z_n)z_n^{t-1}]
&=\lim_{z\to z_n}[(z-z_n)\widetilde{G_{ij}}(z)z^{t-1}]\\
&=-z_n^{t-1}\left(\widetilde{Q_{ij}}(z_n)-\widetilde{Q_{jj}}(z_n)-\delta_{ij}\right)\\
&\times\lim_{z\to z_n}\frac{z-z_n}{(1-z^{-1})\widetilde{Q_{jj}}(z)+P_{j}^\infty}
\end{split}
\end{equation}
Then by applying the l'Hospital's rule, we can get the formula of residues $Res[\widetilde{G_{ij}}(z_n)z_n^{t-1}]$, Eq(\ref{eq_residue}) and Eq.(\ref{eq_coeff}). 

\begin{acknowledgments}
Acknowledgments: This work is supported by ... 
\end{acknowledgments}

\bibliography{DING_2017_03_26_Paper_01_v5.0_Reference}
\end{document}